\documentclass[preprint]{aastex}

\usepackage{natbib}

\shorttitle{VERITAS Observations of 3C\,66A} 
\shortauthors{V. A. Acciari et al. (VERITAS collaboration)}

\begin{document}

\title{VERITAS Observations of a Very High Energy $\gamma$-ray Flare
  from the Blazar 3C\,66A}
\author{
V. A. Acciari\altaffilmark{1},
E. Aliu\altaffilmark{2},
T. Arlen\altaffilmark{3},
M. Beilicke\altaffilmark{4},
W. Benbow\altaffilmark{5},
M. B{\"o}ttcher\altaffilmark{6},
S. M. Bradbury\altaffilmark{7},
J. H. Buckley\altaffilmark{4},
V. Bugaev\altaffilmark{4},
Y. Butt\altaffilmark{8},
K. Byrum\altaffilmark{9},
A. Cannon\altaffilmark{10},
O. Celik\altaffilmark{3},
A. Cesarini\altaffilmark{11},
Y. C. Chow\altaffilmark{3},
L. Ciupik\altaffilmark{12},
P. Cogan\altaffilmark{13},
W. Cui\altaffilmark{14},
M. K. Daniel\altaffilmark{7,\dagger},
R. Dickherber\altaffilmark{4},
T. Ergin\altaffilmark{8},
A. Falcone\altaffilmark{15},
S. J. Fegan\altaffilmark{3},
J. P. Finley\altaffilmark{14},
P. Fortin\altaffilmark{16},
L. Fortson\altaffilmark{12},
A. Furniss\altaffilmark{17},
D. Gall\altaffilmark{14},
K. Gibbs\altaffilmark{5},
G. H. Gillanders\altaffilmark{11},
S. Godambe\altaffilmark{18},
J. Grube\altaffilmark{10},
R. Guenette\altaffilmark{13},
G. Gyuk\altaffilmark{12},
D. Hanna\altaffilmark{13},
E. Hays\altaffilmark{19},
J. Holder\altaffilmark{2},
D. Horan\altaffilmark{20},
C. M. Hui\altaffilmark{18},
T. B. Humensky\altaffilmark{21},
A. Imran\altaffilmark{22},
P. Kaaret\altaffilmark{23},
N. Karlsson\altaffilmark{12},
M. Kertzman\altaffilmark{24},
D. Kieda\altaffilmark{18},
J. Kildea\altaffilmark{5},
A. Konopelko\altaffilmark{25},
H. Krawczynski\altaffilmark{4},
F. Krennrich\altaffilmark{22},
M. J. Lang\altaffilmark{11},
S. LeBohec\altaffilmark{18},
G. Maier\altaffilmark{13},
A. McCann\altaffilmark{13},
M. McCutcheon\altaffilmark{13},
J. Millis\altaffilmark{26},
P. Moriarty\altaffilmark{1},
R. Mukherjee\altaffilmark{16},
T. Nagai\altaffilmark{22},
R. A. Ong\altaffilmark{3},
A. N. Otte\altaffilmark{17},
D. Pandel\altaffilmark{23},
J. S. Perkins\altaffilmark{5,*},
D. Petry\altaffilmark{27},
F. Pizlo\altaffilmark{14},
M. Pohl\altaffilmark{22},
J. Quinn\altaffilmark{10},
K. Ragan\altaffilmark{13},
L. C. Reyes\altaffilmark{28},
P. T. Reynolds\altaffilmark{29},
E. Roache\altaffilmark{5},
H. J. Rose\altaffilmark{7},
M. Schroedter\altaffilmark{22},
G. H. Sembroski\altaffilmark{14},
A. W. Smith\altaffilmark{9},
D. Steele\altaffilmark{12},
S. P. Swordy\altaffilmark{21},
M. Theiling\altaffilmark{5},
J. A. Toner\altaffilmark{11},
A. Varlotta\altaffilmark{14},
V. V. Vassiliev\altaffilmark{3},
R. G. Wagner\altaffilmark{9},
S. P. Wakely\altaffilmark{21},
J. E. Ward\altaffilmark{10},
T. C. Weekes\altaffilmark{5},
A. Weinstein\altaffilmark{3},
D. A. Williams\altaffilmark{17},
S. Wissel\altaffilmark{21},
M. Wood\altaffilmark{3},
B. Zitzer\altaffilmark{14}
}

\altaffiltext{1}{Department of Life and Physical Sciences, Galway-Mayo Institute of Technology, Dublin Road, Galway, Ireland}
\altaffiltext{2}{Department of Physics and Astronomy and the Bartol Research Institute, University of Delaware, Newark, DE 19716, USA}
\altaffiltext{3}{Department of Physics and Astronomy, University of California, Los Angeles, CA 90095, USA}
\altaffiltext{4}{Department of Physics, Washington University, St. Louis, MO 63130, USA}
\altaffiltext{5}{Fred Lawrence Whipple Observatory, Harvard-Smithsonian Center for Astrophysics, Amado, AZ 85645, USA}
\altaffiltext{6}{Astrophysical Institute, Department of Physics and Astronomy, Ohio University, Athens, OH 45701}
\altaffiltext{7}{School of Physics and Astronomy, University of Leeds, Leeds, LS2 9JT, UK}
\altaffiltext{8}{Harvard-Smithsonian Center for Astrophysics, 60 Garden Street, Cambridge, MA 02138, USA}
\altaffiltext{9}{Argonne National Laboratory, 9700 S. Cass Avenue, Argonne, IL 60439, USA}
\altaffiltext{10}{School of Physics, University College Dublin, Belfield, Dublin 4, Ireland}
\altaffiltext{11}{School of Physics, National University of Ireland, Galway, Ireland}
\altaffiltext{12}{Astronomy Department, Adler Planetarium and Astronomy Museum, Chicago, IL 60605, USA}
\altaffiltext{13}{Physics Department, McGill University, Montreal, QC H3A 2T8, Canada}
\altaffiltext{14}{Department of Physics, Purdue University, West Lafayette, IN 47907, USA }
\altaffiltext{15}{Department of Astronomy and Astrophysics, 525 Davey Lab, Pennsylvania State University, University Park, PA 16802, USA}
\altaffiltext{16}{Department of Physics and Astronomy, Barnard College, Columbia University, NY 10027, USA}
\altaffiltext{17}{Santa Cruz Institute for Particle Physics and Department of Physics, University of California, Santa Cruz, CA 95064, USA}
\altaffiltext{18}{Physics Department, University of Utah, Salt Lake City, UT 84112, USA}
\altaffiltext{19}{N.A.S.A./Goddard Space-Flight Center, Code 661, Greenbelt, MD 20771, USA}
\altaffiltext{20}{Laboratoire Leprince-Ringuet, Ecole Polytechnique, CNRS/IN2P3, F-91128 Palaiseau, France}
\altaffiltext{21}{Enrico Fermi Institute, University of Chicago, Chicago, IL 60637, USA}
\altaffiltext{22}{Department of Physics and Astronomy, Iowa State University, Ames, IA 50011, USA}
\altaffiltext{23}{Department of Physics and Astronomy, University of Iowa, Van Allen Hall, Iowa City, IA 52242, USA}
\altaffiltext{24}{Department of Physics and Astronomy, DePauw University, Greencastle, IN 46135-0037, USA}
\altaffiltext{25}{Department of Physics, Pittsburg State University, 1701 South Broadway, Pittsburg, KS 66762, USA}
\altaffiltext{26}{Department of Physics, Anderson University, 1100 East 5th Street, Anderson, IN 46012}
\altaffiltext{27}{European Southern Observatory, Karl-Schwarzchild-Strasse 2, 85748 Garching, Germany}
\altaffiltext{28}{Kavli Institute for Cosmological Physics, University of Chicago, Chicago, IL 60637, USA}
\altaffiltext{29}{Department of Applied Physics and Instumentation,
  Cork Institute of Technology, Bishopstown, Cork, Ireland}

\altaffiltext{$\dagger$}{Now at: Department of Physics, Durham University, South Road, Durham, DH1 3LE, U.K.}
\altaffiltext{*}{Corresponding author: jperkins@cfa.harvard.edu}

\begin{abstract}
  The intermediate-frequency peaked BL Lacertae (IBL) object 3C\,66A
  is detected during 2007 - 2008 in VHE (very high energy: E $>$ 100
  GeV) $\gamma$-rays with the VERITAS stereoscopic array of imaging
  atmospheric Cherenkov telescopes.  An excess of 1791 events is
  detected, corresponding to a significance of 21.2 standard
  deviations ($\sigma$), in these observations (32.8 hours live time).
  The observed integral flux above 200 GeV is 6\% of the Crab Nebula's
  flux and shows evidence for variability on the time-scale of days.
  The measured energy spectrum is characterized by a soft power law
  with photon index $\Gamma=4.1 \pm 0.4_{stat} \pm 0.6_{sys}$. The
  radio galaxy 3C\,66B is excluded as a possible source of the VHE
  emission.
\end{abstract}


\section{Introduction}

\citet{Wills:1974eu} first identified 3C\,66A as a QSO using optical
observations. It was subsequently classified as a BL Lac object based
on its significant optical and X-ray variability
\citep{Maccagni:1987qe}. BL Lac objects are characterized by a
double-humped spectral energy distribution (SED) and are further
classified according to the location of the lower energy hump, usually
interpreted as synchrotron emission from relativistic
electrons. \cite{Perri:2003mi} locate the synchrotron peak between
$10^{15}$ and $10^{16}$ Hz. Therefore 3C\,66A is classified as an
intermediate-frequency peaked BL Lac (IBL). To date, the majority of
BL Lacs detected at VHE (very high energy: E $>$ 100 GeV) are HBLs
(high-frequency peaked BL Lacs). Only one other IBL, W Comae, has been
detected above 100 GeV \citep{V.A.Acciari:2008fy}.

During states of high flux, continuum emission from the jet is
dominant and overshadows the few emission lines from the rest of the
galaxy. Blazars have few, if any, detectable emission lines, which
makes determining the redshift difficult even under the best
conditions. Based on a single line, interpreted as Mg II, 3C\,66A was
determined to be at a redshift of $z = 0.444$ \citep{Miller:1978wq}.
In addition to this, \citet{Lanzetta:1993ao} identified a weak
Ly-alpha line corroborating these results.  Since both measurements
rely on a single line, the redshift of this BL Lac is considered
uncertain. \citet{Finke:2008zm} recently derived a lower limit of $z =
0.096$.  The determination of the redshift is crucial to understanding
this source at VHE energies due to the effect of the extragalactic
background light (EBL) \citep{Hauser:2001jt}.  VHE $\gamma$ rays are
absorbed via pair production interactions with the infrared component
of the EBL ($\gamma_{vhe}\gamma_{ebl} \rightarrow e^+e^-$)
\citep{Gould:1967nx}. At VHE energies, this absorption causes a
decrease in the observed flux and a softening of the observed
spectrum.  One can calculate an optical depth ($\tau(z,E)$) based on
an EBL density model, the redshift and the $\gamma$-ray energy.  The
optical depth can be used to calculate the flux corrected for
extragalactic absorption from the observed flux at a given energy
($F_{int} = e^{\tau(z,E)}F_{obs}$).  Without an accurate measure of
the redshift, an accurate photon spectrum intrinsic to the blazar
cannot be calculated or modeled.

The EGRET source 3EG J0222+4253, detected at an integral flux between
$(12.1 \pm 3.9) \times 10^{-8}$ and $(25.3 \pm 5.8) \times 10^{-8}$
photons cm$^{-2}$ s$^{-1}$, is associated with 3C\,66A
\citep{Hartman:1999fu}.  Measurements of the spectrum indicated that
the spectral index of $2.01 \pm 0.14$ was influenced by the nearby
pulsar PSR 0218+42, which is also inside the EGRET error box.  A
detailed study of the energy-dependent position of the EGRET source
shows that the highest energy photons are coming from the BL Lac and
thus the spectrum is thought to continue out to the VHE band
\citep{Kuiper:2000dk}.  A re-analysis of the EGRET data by
\citet{Nandikotkur:2007jt} finds a harder spectral index of
$1.95\pm0.14$ and a flux above 100 MeV of $(17.7\pm2.8) \times
10^{-8}$ photons cm$^{-2}$ s$^{-1}$. Recently, the Fermi Gamma-ray
Space Telescope also reported a detection of 3C\,66A, contemporaneous
with VHE data taken by VERITAS in the 2008-2009 season, at a higher
flux than previously reported by EGRET \citep{tosti:2008}.

There have been several attempts to detect 3C\,66A in VHE $\gamma$
rays.  The Crimean Astrophysical Observatory reported a 5.1 $\sigma$
detection above 900 GeV at an average integral flux of $2.4 \times
10^{-11} $ cm$^{-2}$ s$^{-1}$ \citep{Stepanyan:2002ss}.  Additionally,
both the Whipple 10~m telescope \citep{Horan:2004yo} and HEGRA
telescope array \citep{Aharonian:2000qc} observed 3C\,66A and reported
upper limits of $< 0.35 \times 10^{-11}$cm$^{-2}$s$^{-1}$ (99.9\%
confidence; above 350 GeV) and $< 1.4\times 10^{-11}$cm$^{-2}$s$^{-1}$
(99\% confidence; above 630 GeV), respectively.  STACEE reported
several upper limits between $< 1.0$ and $< 1.8 \times 10^{-10}$
cm$^{-2}$ s$^{-1}$ above 150 - 200 GeV depending on the source
spectrum \citep{Bramel:2005nx}.  Recently, MAGIC reported a
$5.4~\sigma$ detection of VHE emission above 150 GeV from observations
in September to December 2007 coincident with 3C\,66B.  They exclude
3C\,66A as the source of the VHE emission at an 85\% confidence level
\citep{Aliu:2009ly}.

\section{VERITAS Detector \& Observations}

The VERITAS detector is an array of four 12~m diameter imaging
atmospheric Cherenkov telescopes located in southern Arizona
\citep{Weekes:2002pi}.  Designed to detect emission from astrophysical
objects in the energy range from 100 GeV to greater than 30 TeV,
VERITAS has an energy resolution of $\sim$15\% and an angular
resolution (68\% containment) of $\sim0.1^\circ$ per event.  A source
with a flux of 1\% of the Crab Nebula is detected in $\sim 50$ hours
of observations while a 5\% Crab Nebula flux source is detected in
$\sim 2.5$ hours.  The field of view of the VERITAS telescopes is
$3.5^\circ$.  For more details on the VERITAS instrument and
technique, see \citet{Holder:2008}.

VERITAS observed 3C\,66A for 14 hours from September 2007 through
January 2008 (hereafter, the 2007-2008 season).  From September
through November 2008 (hereafter, the 2008-2009 season), a further 46
hours of data were taken. In total, 180 twenty-minute exposures were
made, where 109 exposures passed selection criteria which remove data
with poor weather (based on infrared sky-temperature measurements) or
with hardware-related problems. Data collection frequently occurred
during poor weather conditions causing the lower selection throughput
demonstrated here. In total, the 2007-2008 season resulted in 4.7
hours live time and the recent 2008-2009 season produced 28.1 hours
live time. The average zenith angle was $17.3^\circ$. All data were
taken on moonless nights in ``wobble'' mode where the telescopes are
pointed away from the source by $\pm 0.5^\circ$ to allow for
simultaneous background estimation \citep{Berge:2007ud}.

\section{Analysis Methods}
   
Prior to event selection and background subtraction, the shower images
are calibrated and cleaned as described in \cite{Cogan:2006} and
\cite{Daniel:2007kx}.  Several noise-reducing event-selection cuts are
made at this point, including rejecting those events where only the
two closest-spaced telescopes participated in the trigger. Following
the calibration and cleaning of the data, the events are parametrized
using a moment analysis \citep{Hillas:1985ta}.  From this moment
analysis, scaled parameters are calculated and used for event
selection \citep{Aharonian:1997rm,Krawczynski:2006ts}. The event
selection cuts are optimized {\it a priori} using data taken on the
Crab Nebula, scaling the background and excess rates to account for a
weaker source. These selection criteria are termed the ``standard
cuts''. Since 3C\,66A is possibly very distant and the observed
spectrum is expected to be soft, a modified ``soft cuts'' applies a
further {\it a priori} optimization of increasing the $\theta^2$ cut
(the angular distance squared from the position of 3C\,66A and the
reconstructed shower direction) and decreasing the size cut (the
number of photo-electrons in an event) for sources with a soft
spectrum (see Table \ref{tab:cuts}).  Unless stated otherwise, the
``soft cuts'' were used to generate the results presented in this
paper.

\begin{table}
\begin{center}
  \caption{Selection cuts applied to the data for the soft cuts as
    well as the standard cuts\tablenotemark{a}. Also shown are the
    results of the VERITAS observations of 3C\,66A using both the
    standard analysis cuts and soft spectrum cuts.\label{tab:cuts}}
\begin{tabular}{ccccccccc}
\tableline\tableline
Cuts & $\theta^2$ & Size\tablenotemark{b} & E$_{th}$\tablenotemark{c} & On\tablenotemark{d} & Off\tablenotemark{e} & Alpha\tablenotemark{f} & Excess\tablenotemark{g} & Sig. \\
     & [deg$^2$] & [DC] & [GeV]  & & & & & [$\sigma$] \\
\tableline
Soft & $< 0.020$ & $> 200$ & 120 & 7257 & 31201 & 0.1752 & 1791 & 21.1 \\
Std. & $< 0.013$ & $> 400$ & 170 & 1258 & 5282  & 0.1400 & 518 & 16.0 \\
\tableline
\end{tabular}

\tablenotetext{a}{All other cuts outside of those shown here are the same}
\tablenotetext{b}{The photo-electron to digital count ratio is
  approximately four}
\tablenotetext{c}{Post-cuts energy threshold derived using a spectral
  index of 4.1} 
\tablenotetext{d}{Number of on-source events passing cuts}
\tablenotetext{e}{Number of off-source events passing cuts}
\tablenotetext{f}{Normalization for the off-source events}
\tablenotetext{g}{Observed Excess}

\end{center}
\end{table}

The reflected-region model \citep{Berge:2007ud} is used for background
subtraction. The total number of events in the on-source region is
then compared to the total number of events in the more numerous
off-source regions, scaled by the ratio (called $\alpha$) of the solid
angles, to produce a final excess.

\section{VERITAS Results}

\begin{figure}
\plotone{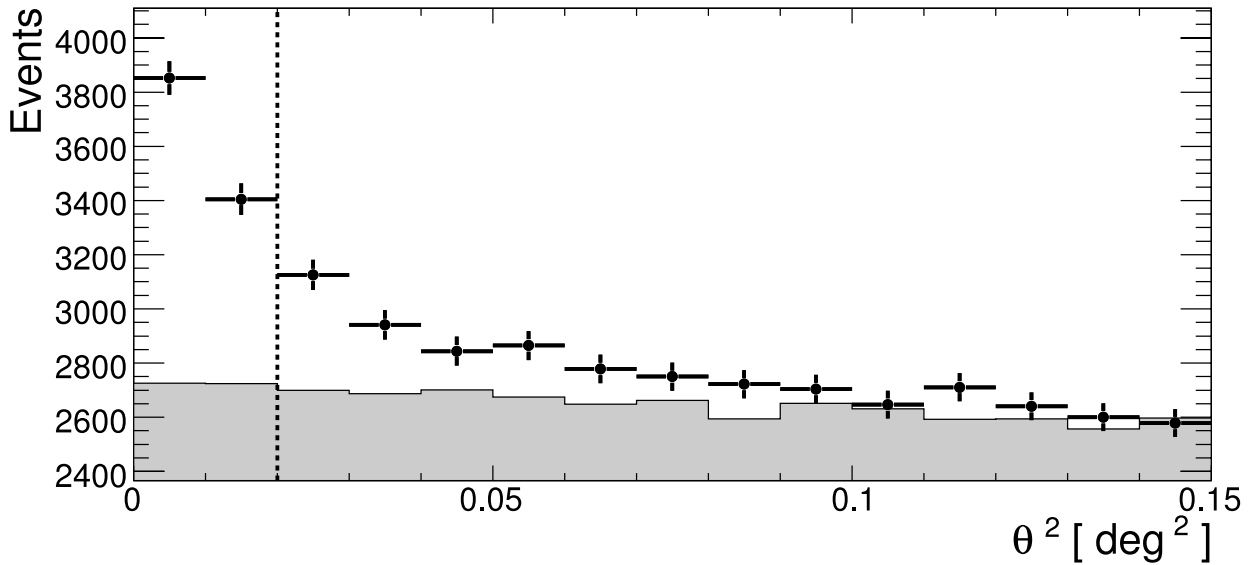}
\caption{\label{fig:theta2} The distribution of $\theta^2$ for
  on-source events (points) and normalized off-source events (shaded
  region) from observations of 3C\,66A.  The dashed line represents the
  cut on $\theta^2$ applied to the data.}
\end{figure}

\begin{figure}
\plotone{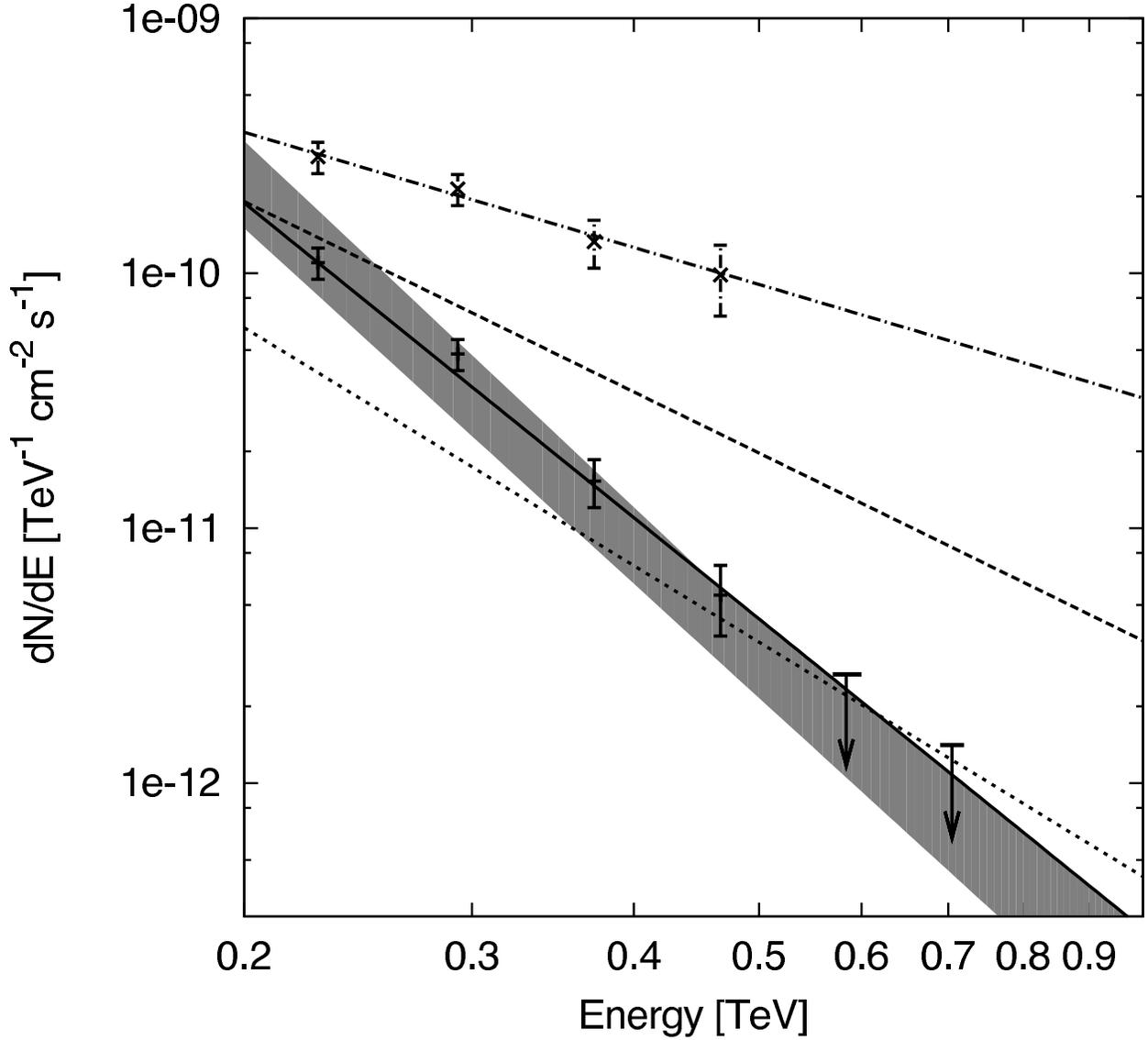}
\caption{\label{fig:spectrum} The energy spectrum of 3C\,66A shown as
  solid points.  The spectrum is well fitted by a power law with index
  $\Gamma=4.1 \pm 0.4_{stat} \pm 0.6_{sys}$ (solid line). The shaded
  area outlines the systematic error in the spectral index. Using the
  models of \citet{Franceschini:2008il} and assuming a redshift of
  $z=0.444$, the de-absorbed spectral index is calculated to be $1.5
  \pm 0.4$ showing that the very steep measured spectrum could be due
  to the distance of 3C\,66A.  This de-absorbed spectrum is shown as a
  dashed-dotted line and points.  The MAGIC spectrum with index
  $\Gamma=3.1$ from \citet{Aliu:2009ly} is shown as a dotted line.
  The Crab Nebula's spectrum divided by 10 is also shown for
  comparison (dashed line).}
\end{figure}

\begin{figure}
\plotone{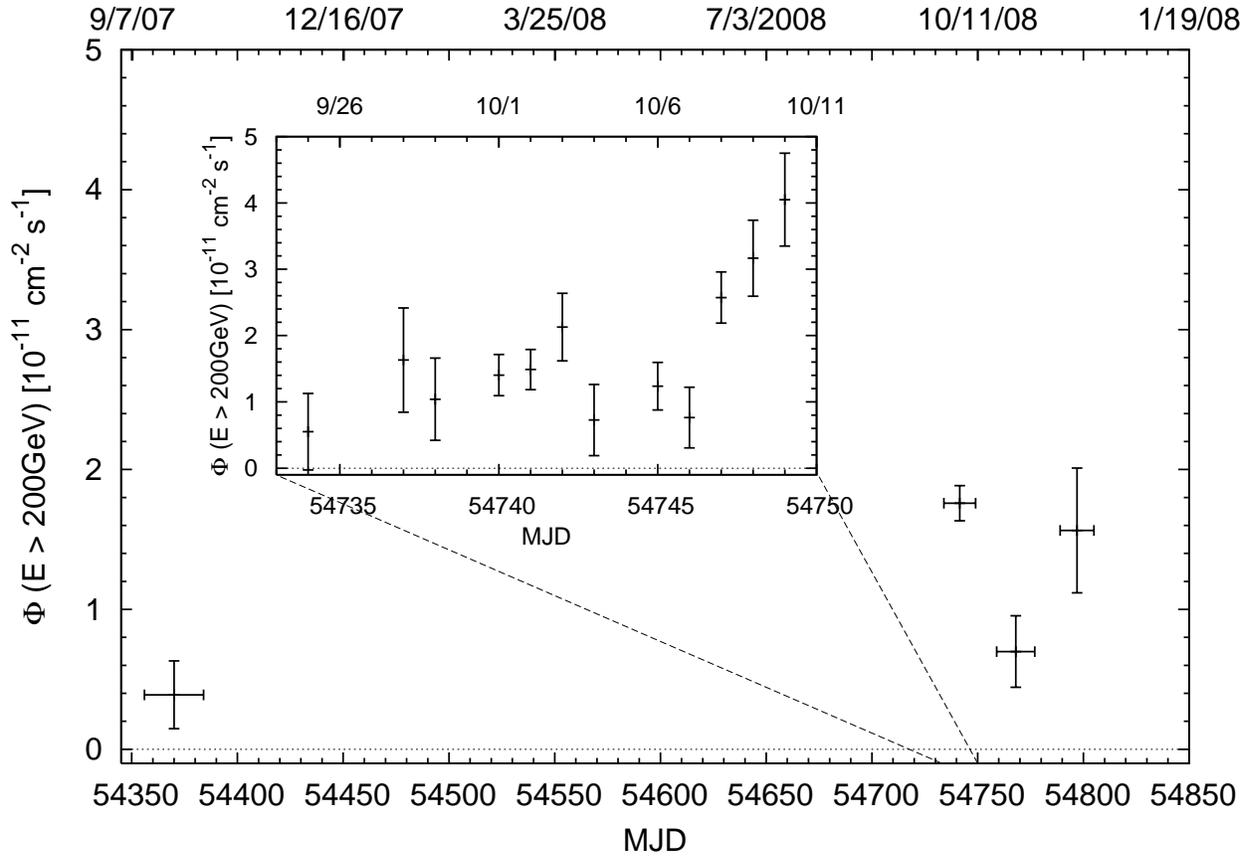}
\caption{\label{fig:2008lc} Light curve binned by dark period (the
  time between two full moons) for the full data set from 3C\,66A.
  These data indicate night-by-night variability for the second dark
  period (MJD 54734 through 54749) but not within any of the other
  dark periods.  The inset details this dark period binned by night
  when the flare occurred.}
\end{figure}

\begin{figure}
\plotone{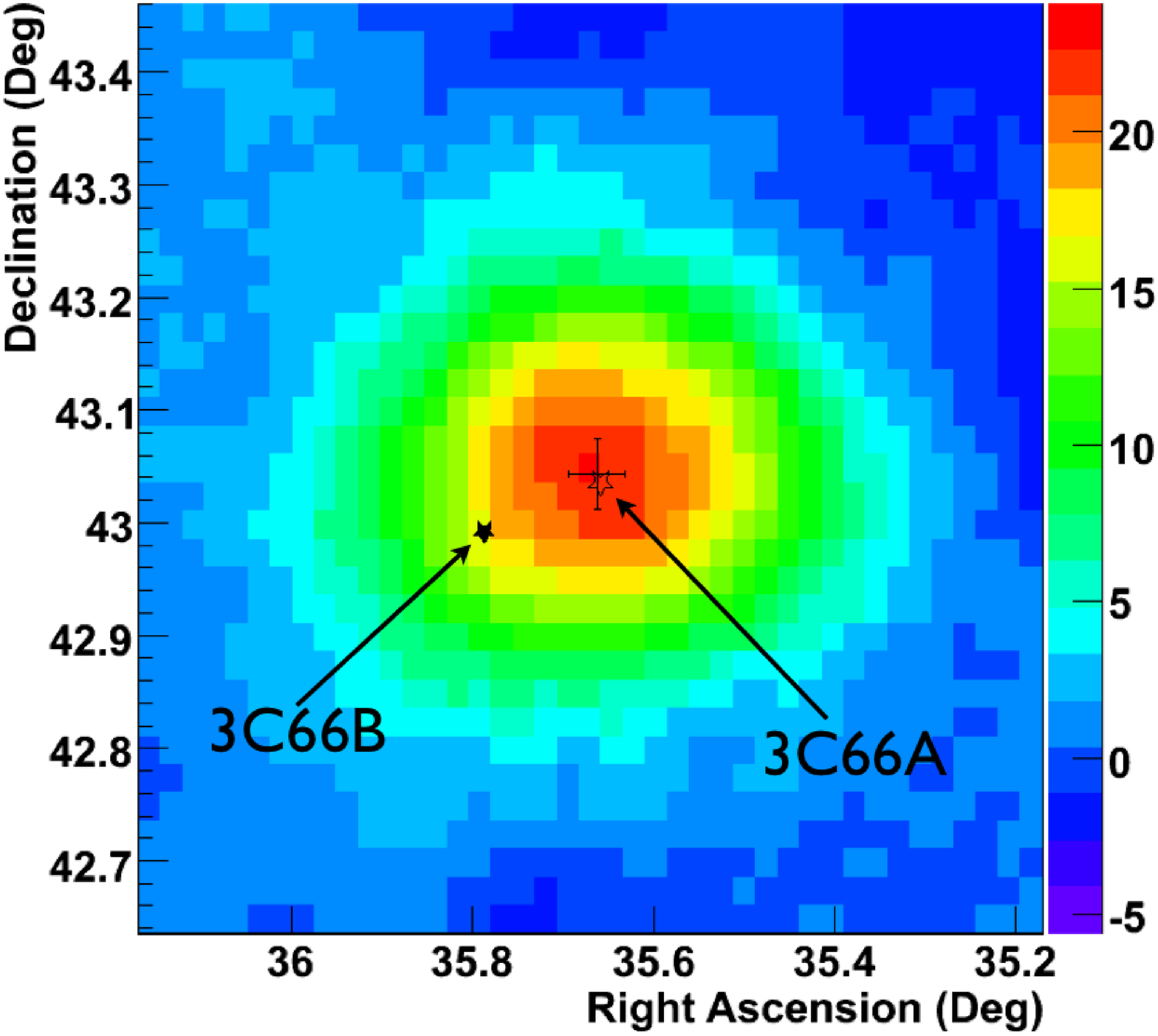}
\caption{\label{fig:location} Smoothed significance map of 3C\,66A.
  The location of 3C\,66A is shown as an open star and 3C\,66B as the
  closed star.  The cross is the fit to the excess VHE emission
  resulting in a localization of 2h 22m 41.6s $\pm$ 1.7s $\pm$ 6.0s,
  $43^o$ 02' 35.5'' $\pm$ 21'' $\pm$ 1'30''.  These data strongly
  favor 3C\,66A as the source of the $\gamma$-ray emission at a
  significance level of $4.3~\sigma$. Note that the bins are highly
  correlated due to an integration over angular space.}
\end{figure}

An excess of 1791 events is observed from the direction of 3C\,66A
(7257 on events, 31201 off events with an off-source normalization
ratio $\alpha$ of 0.1752).  The excess corresponds to a statistical
significance of 21.1 $\sigma$ using Equation 17 from
\citet{Li:1983pb}.  The distribution of $\theta^2$ is shown in Figure
\ref{fig:theta2}. The shape of the excess is consistent with that
expected from a point source (68\% containment in $0.16^\circ$). Since
lower-energy events are more poorly reconstructed and this data set is
dominated by low-energy events, the $\theta^2$ distribution is wider
than that of a harder-spectrum source. Table \ref{tab:cuts} details
the results using soft cuts as well as standard cuts.  Note that the
use of two different sets of cuts has an associated trials factor of
two, which has negligible impact on the significance. 

Using the soft cuts, the differential energy spectrum over the energy
range $\sim200$ GeV to $\sim500$ GeV is determined and is shown in
Figure 2.  The best fit of a power-law to these data yields an index
$\Gamma=4.1 \pm 0.4_{stat} \pm 0.6_{sys}$ with a chi-square of 1.94
for 2 degrees of freedom. An alternative analysis chain confirms this
very soft spectrum. Note that there were 1431 excess events detected
during the flaring period from MJD 54740 through MJD 54749 which
accounts for 80\% of the total.  Thus, while the spectrum calculated
here is for the full data set, it is dominated by the flare.  Assuming
this power-law spectrum, the observed integral flux for the full data
set above 200 GeV is $(1.3 \pm 0.1) \times 10^{-11}$ cm$^{-2}$
s$^{-1}$ ($6\%$ of the Crab Nebula's flux). By comparison, the
2007-2008 season yielded a significance of $2.6~\sigma$ at a lower
average flux above 200 GeV of $(3.9 \pm 1.6) \times 10^{-12}$ photons
cm$^{-2}$ s$^{-1}$, which is 26\% of the flux seen in
2008-2009. Figure \ref{fig:2008lc} shows the integral flux above 200
GeV from 3C\,66A for each dark period (the time between two full
moons). The highest flux seen from 3C 66A occurred on MJD
54749. Significant variability is seen only during the dark period
spanning September 25 through October 10 (shown in the inset in Figure
\ref{fig:2008lc}), with a chi-squared probability of 0.009\% for a fit
to a constant flux. No statistically significant evidence for
variability is seen within any of the individual nights.  Fits of a
constant to the nightly flux in any other dark period do not yield a
chi-squared probability less than 10\%.

The radio galaxy 3C\,66B lies in the same field of view as 3C\,66A at
a separation of $0.12^\circ$ and is also a plausible source of VHE
radiation \citep{Tavecchio:2008nx}. With the recent detection of VHE
emission from the 3C\,66A/B region by MAGIC \citep{Aliu:2009ly}
favoring 3C\,66B as the source and excluding 3C\,66A at an 85\%
confidence level, it is important to determine which of these objects
is the source of the emission reported here. Thus, a 2-dimensional
Gaussian shape was fit to the uncorrelated excess of $\gamma$ rays,
yielding a position of 2h 22m 41.6s $\pm$ 1.7s, $43^o$ 02' 35.5''
$\pm$ 21'', with an additional systematic angular uncertainty of 90''.
The systematic error has been confirmed via optical pointing monitors
which are mounted to each telescope.  This rules out 3C\,66B as the
source of the VHE emission reported here at a significance level of
$4.3~\sigma$. 3C\,66A lies $0.01^\circ$ from the fit position while
3C\,66B lies $0.13^\circ$ away and the total error on the fit is
$0.03^\circ$ (see Figure \ref{fig:location}).  In addition to fitting
the full data set, fits were made to the data divided into high
($>300$ GeV) and low ($<300$ GeV) energy bands under the assumption
that the high-energy emission might originate from 3C\,66B while the
low energy emission might come from 3C\,66A
\cite[see][]{Tavecchio:2008nx}.  The fit to the position did not
deviate from the measurement using the full data set. Correlated
variability studies utilizing optical, VHE and HE (30 MeV - 100 GeV)
bands are underway to verify 3C\,66A as the source of VHE $\gamma$
rays and will be the subject of a future paper.  Further restricting
the VERITAS data to observations contemporaneous with MAGIC in
September to December 2007, we calculate an upper limit, assuming the
reported MAGIC spectrum of $\Gamma = 3.10$, on the flux above 300 GeV
from 3C\,66B to be $1.8 \times 10^{-12}$ photons cm$^{-2}$ s$^{-1}$ at
the 99\% confidence level based on $\sim5$ hours of data. MAGIC
reported an integral flux based on a $\sim50$ hour exposure above 150
GeV of $(7.3 \pm 1.5) \times 10^{-12}$ photons cm$^{-2}$ s$^{-1}$ for
their full data set (approximately $1.7 \times 10^{-12}$ photons
cm$^{-2}$ s$^{-1}$ above 300 GeV). Unfortunately, it is not possible
to calculate a spectrum from the 2007-2008 season data due to low
statistics.  Although VERITAS is more sensitive ($\sim2x$) than MAGIC
the brief ($\sim5$ hour) exposure on the 3C\,66A/B region in 2007 does
not enable a clear determination of which object was the source of VHE
emission in 2007.  However, based on the MAGIC flux, VERITAS expects
$\sim700$ excess events at the location of 3C\,66B in the full data
set, whereas only $\sim300$ are detected and the latter is consistent
with expectations for spill over from 3C\,66A due to the VERITAS point
spread function. Therefore, if the MAGIC claims of VHE emission from
3C\,66B are correct, it must have been considerably brighter in 2007
than 2008, and similarly 3C 66A must have been considerably brighter
in 2008 than 2007.

\section{Summary \& Conclusion}

VERITAS has observed the IBL 3C\,66A for a total of 32.8 hours
good-quality live time from September 2007 through November 2008,
resulting in the detection of VHE $\gamma$ rays with a statistical
significance of 21.1 $\sigma$.  The average integral flux above 200
GeV is $(1.3 \pm 0.1) \times 10^{-11}$ cm$^{-2}$ s$^{-1}$ ($6\%$ of
the Crab Nebula's flux).  The differential energy spectrum is well fit
by a soft power law with index $\Gamma=4.1 \pm 0.4_{stat} \pm
0.6_{sys}$ between 200 and 500 GeV.

It is thought that the redshift of 3C\,66A is $z=0.444$ but this
measurement is based upon a single, poorly detected line
\citep{Miller:1978wq}.  A definitive measurement of the redshift is
needed to determine the intrinsic spectrum of 3C\,66A, corrected for
EBL absorption.  The extreme distance of 3C\,66A, if true, will allow
modelers to probe the evolution of the EBL with redshift.  Assuming
the current redshift measurement of $z=0.444$, we calculate a
de-absorbed spectrum based on the EBL models of
\citet{Franceschini:2008il} which are based upon recent measurements
from the optical to the sub-millimeter.  The original spectrum along
with the corrected spectrum can be seen in Figure \ref{fig:spectrum}.
While this is not a definitive calculation of the intrinsic spectrum,
due to the uncertainties in the redshift measurement and in the
modeling of the EBL, it illustrates that the steepness of the measured
spectrum could be due to the distance of 3C\,66A.

The initial announcement of a detection of VHE emission from 3C\,66A
\citep{swordy:2008} prompted several other groups and instruments to
also observe this object \citep{tosti:2008, Larionov:2008}.  In
addition, the Fermi Gamma-ray Space Telescope detected 3C\,66A at a
level higher than that reported by EGRET.  The Swift observatory also
monitored 3C\,66A over this time period, in the X-ray and UV bands.  A
future paper by the Fermi collaboration, VERITAS collaboration and
multi-wavelength partners will describe these results and provide
details on correlated variability as well as a broadband spectral
energy distribution for the 2008 data set.

\acknowledgements

This research was supported by grants from the U.S. Department of
Energy, the U.S. National Science Foundation and the Smithsonian
Institution, by NSERC in Canada, by Science Foundation Ireland and by
STFC in the UK.

\bibliographystyle{apj}

\end{document}